\documentclass[letterpaper]{article} %
\usepackage{aaai2026}  %
\usepackage{times}  %
\usepackage{helvet}  %
\usepackage{courier}  %
\usepackage[hyphens]{url}  %
\usepackage{graphicx} %
\usepackage{natbib}  %
\usepackage{caption} %
\usepackage{algorithm}
\usepackage{algorithmic}

\usepackage{newfloat}
\usepackage{listings}
\DeclareCaptionStyle{ruled}{labelfont=normalfont,labelsep=colon,strut=off} %
\floatstyle{ruled}
\newfloat{listing}{tb}{lst}{}
\floatname{listing}{Listing}

\usepackage[utf8]{inputenc}
\usepackage{%
amssymb,amsthm,amsfonts}
\usepackage{cleveref}
\usepackage{thm-restate}
\usepackage{enumitem}
\usepackage{subcaption}
\usepackage{placeins}

\usepackage{array}
\newcolumntype{C}[1]{>{\centering\arraybackslash\hyphenpenalty=10000\exhyphenpenalty=10000}p{#1}}

\usepackage[textsize=tiny]{todonotes}
\usepackage{csquotes}
\newcommand{\Comments}{1}
\newcommand{\mynote}[2]{\ifnum\Comments=1\textcolor{#1}{#2}\fi}
\newcommand{\mytodo}[2]{\ifnum\Comments=1%
	\todo[linecolor=#1!80!black,backgroundcolor=#1,bordercolor=#1!80!black]{#2}\fi}

\usepackage[group-separator={,}]{siunitx}
\title{Algorithmic Gender Prediction Is Illegitimate, But Gender Imputation Can Yield Valid Measurements}

\author{Evan Dong and Angelina Wang}
\affiliations{
    Cornell University\\
    edong.cs@cornell.edu, angelina.wang@cornell.edu
}

\begin{document}

\maketitle

\begin{abstract}
    Machine learning ethics researchers, critical HCI scholars, and interdisciplinary academics have argued that algorithmically predicting gender is \textit{wrong}. At the same time, others in the same research communities rely on predicted gender labels to study gender disparities and develop algorithmic fairness techniques. Taking both of these perspectives in good faith pushes us to believe that algorithmic gender prediction simultaneously cannot be trusted and at the same time can identify genuine gender disparities. How do we reconcile these two seemingly contradictory intuitions? We draw a distinction between two questions: whether algorithmic gender prediction is wrong, being \textit{illegitimate} and thereby contributing to harm; and whether algorithmic gender prediction is wrong, being \textit{invalid} and thereby producing unusable measurements. Our analysis translates different arguments against gender prediction into these terms of legitimacy and validity and shows how gender imputation applied for fairness purposes can be both illegitimate and still yield valid disparity measurements. We clarify this bind by drawing upon transfeminist literature to distinguish two forms of sexism: sexism that targets women and femininity, and sexism that targets transgender and nonbinary people. While gender imputation can produce valid measurements for the former, it is illegitimate and harmful for the latter. 
    We argue that practitioners should deploy gender imputation only when it would achieve anti-discrimination benefits that cannot be achieved through other reasonable means, while harms are minimized to the extent possible. We examine how this tension plays out through three case studies: auditing gender bias in generative image models, measuring gender disparities in film, and imputing gender from personal names. By disentangling both questions of legitimacy from questions of validity, as well as these two different forms of sexism, we show how debates over gender prediction have often conflated distinct concerns, obscuring both the settings in which gender imputation can support fairness efforts and the harms towards transgender and nonbinary people that it fundamentally cannot capture. We conclude by recommending the development of more inclusive methods that address all kinds of sexism.

\end{abstract}

\section{Introduction}

Assessing algorithmic discrimination generally requires access to demographic data. Consider that quantifying the U.S. gender pay gap in 2024 --- that women earned 85\% of what men earned --- requires data on workers’ genders~\citep{fry2025gender}. However, demographic data is not always available. For instance, in the U.S., many financial institutions and government agencies such as the Internal Revenue Service and Patent and Trademark Office are often not legally permitted to collect demographic data such as race and sex~\citep{bogen2020awareness,king2023privacy,kumar2022equalizing}. Similarly, users may be reluctant to disclose their demographic data for privacy reasons, especially to large institutions they may not trust~\citep{bogen2020awareness,king2023privacy,andrus2021we}. In these cases, this demographic data is often \textit{imputed} (i.e., filled in with estimated values via a predictive method); for example, by predicting gender from images or given names, or predicting race from ZIP codes and surnames. Many transgender\footnote{``Trans'' is a common abbreviation of ``transgender,'' especially when used as a modifying adjective,  e.g., ``trans studies'' or ``trans philosophy,'' or a prefix, e.g., ``trans-exclusionary,'' ``transphobia.'' Like \citet{ovalle2023m}, we refer readers to the community-run 
Nonbinary wiki's English glossary of gender terms. For convenience, we have transcribed definitions of terms that appear in this paper into App.~\ref{sec:queer-terms}.}
and critical scholars have opposed this practice when it comes to gender, arguing against prediction on two main grounds: first, the direct harm of misgendering individuals; and second, the problematic premises it encodes, which reify gender as binary, static, and predictable from a face or name. 

In this work, we focus specifically on the case of predicting gender, rather than other demographics such as race or age. While predicting other demographic attributes faces the same bind of label need and premise critique, and many of our arguments may apply,
focusing on gender enables us to engage concretely with the specific critiques posed by, e.g., transgender and nonbinary\footnote{Our language in this paper focuses on transgender and nonbinary people. We use these terms with the understanding that, they cover very heterogeneous experiences, and that others under different labels, such as gender non-conforming and genderqueer, are often similarly marginalized, part of similar communities, and hold similar concerns.} advocates, and the gender-specific harms of this premise~\citep{smith1983approaches,reid1988racism,wang2025identities}.

The evaluation of gender disparity is thus caught between the desire to detect situations in need of change and the desire to dismantle harmful conceptions of static gender. This bind is reflected in conflicting moral intuitions in day-to-day life: consider that many women and nonbinary researchers have experienced the real harms of feeling like the only gender minority when walking into a room full of perceived men at a computer science conference, even without knowing the actual gender identities of those present.
Trans-inclusive principles show us that assuming another person's gender can be wrong; however, these principles do not exist for the sake of invalidating common experiences of gender discrimination. 

We navigate this tension by drawing two key distinctions that have previously muddled this discussion. First, we distinguish legitimacy (whether a prediction normatively deserves given social authority) from validity (whether a prediction descriptively measures a concept well). Second, we draw on Julia Serano's analysis of trans-misogyny~\citep{serano2007whipping} to distinguish traditional sexism (discrimination against women and femininity) and oppositional sexism (discrimination against gender deviance and non-normativity).
We argue that these different forms of sexist discrimination are based on different aspects of gender, such as gender identity and perceived gender, that have different legitimacy and validity implications. This allows us to concretely describe which dimensions of sexism that imputation is more suited to measure (i.e., traditional sexism), while also highlighting the dimensions of sexism that are often ignored or actively perpetrated (i.e., oppositional sexism) when overly focusing on the former. %
This allows us to reconcile this bind and argue that \textit{gender imputation can lead to valid measurement of disparities for traditional sexism, even though its premise is illegitimate and causes harm with respect to oppositional sexism}. We emphasize that this illegitimacy is a serious concern, especially in a time where trans rights are under attack and transgender people are facing intense persecution; our work seeks to navigate this tension without discounting this fact, and in fact to critique works using gender imputation that relegate this harm to a mere acknowledgment or footnote~\citep{devinney2022theories}. As much as we seek to defend imputation for traditional sexism, we also call for action to address the neglect of oppositional sexism.

We lay out our argument in detail before discussing recommendations and case studies.
First, in Sec.~\ref{sec:background} we provide background on the imputation of demographic data and differentiate imputation as a special case of prediction. Next, in Sec.~\ref{sec:illegitimate} we examine existing critiques of gender prediction, and analyze which characteristics affect the legitimacy of the prediction. By drawing on trans scholarship, we highlight a range of applications, prediction targets, and gender concepts, demonstrating that all gender prediction is to some degree illegitimate. Then, in Sec.~\ref{sec:valid} we outline how these reasons for illegitimacy affect (or not) the validity of measurements that are relevant to measuring gender discrimination, with some limitations, and in Sec.~\ref{sec:recommendations} provide concrete recommendations towards better practices in gender imputation. In Sec.~\ref{sec:case-studies} we work through the relevance of different factors in model design and downstream application, sharing three case studies on gender imputation in generative models, film, and personal names, to demonstrate how our argument plays out in practice.
We do not mean to legitimize gender prediction as it is often practiced, which causes severe harm to transgender and nonbinary people --- harm that can and should be minimized. We provide recommendations for such harm minimization, and acknowledge only that, in cases where no alternative methods exist, there are benefits of measuring traditional gender discrimination that should not be ignored. In this work, our core contributions are to: (a) work through the bind of algorithmic gender prediction being illegitimate but sometimes valid, (b) distinguish the suitability of imputation for addressing traditional sexism from oppositional sexism, and (c) offer recommendations to minimize harm, maximize benefit, and guide practitioners and researchers on navigating this bind across different contexts.

\section{Background}
\label{sec:background}

\subsection{Demographic Data Imputation}

Demographic data is often unavailable, for several reasons, including legal barriers to collection. For example, in the United States, the Equal Credit Opportunity Act originally restricted lenders' ability to collect demographic data as a way to \textit{prevent} intentional discrimination by malicious decision-makers. Similarly, France, along with many other European countries, does not collect racial statistics or codify racial categories as part of a policy of race-blindness~\citep{simon2015choice,escafre2011ethnic}. In other cases where data is collected, sharing data between different government agencies may be limited for privacy reasons~\citep{king2023privacy}. Privacy may also be particularly relevant to more sensitive attributes; detecting discrimination on the basis of sexual orientation requires collecting information that some may be (reasonably) hesitant to disclose, so optional data collection suffers from selection bias~\citep{tomasev2021fairness}. As a result, in many practical settings, disparity measurement is hampered by a lack of demographic data. 

A substantial literature in algorithmic fairness and disparity auditing thus examines the issue of fairness or disparity estimation without fully known labels. However, without any assumptions or demographic information at all, problems of algorithmic fairness are largely intractable. Instead, researchers assume noisy, imperfect, or partially available estimates of demographic data. In practice, these estimates often come from predictive models. %
These estimates have been studied both for their statistical properties and downstream fairness implications~\citep{chen2019fairness, dong2025addressing} as well as used for the explicit purposes of assessing disparities~\citep{chen2019fairness, mccartan2023estimating,kwegyir2024observing} and training debiased machine learning models~\citep{wang2020robust, mccartan2023estimating, elzayn2024estimating}. Notably, Bayesian Improved Surname Geocoding (BISG)~\citep{elliott2009using}, a naive Bayes model of race using Census name and location data, has been used to motivate real, anti-discrimination regulation: in 2013, the United States Consumer Financial Protection Bureau (CFPB) used imputation to detect racial disparities in auto lending~\citep{bureau2014using}. Other scholars have used parental country of origin in the French context as a proxy for race and ethnicity~\citep{govind2025racial}.

\subsection{Distinguishing Imputation and Prediction} \label{sec:distinctions}

We have not yet differentiated gender prediction and gender imputation. Here, we present our definitions, where gender imputation is a subset of gender prediction. 

We define \textit{gender prediction} as inferring gender from input(s). In machine learning terms, where a predictive model, ($f$), is characterized by $f(X) = Y$, prediction treats gender as $Y$, assuming that $Y$ is not known a priori.
We define \textit{gender imputation} as a subset of prediction, distinguished by its \textit{application} and \textit{interpretation}. Prediction falls under the category of imputation if its application is a specific, limited end, namely to audit or describe existing structures or processes, and its interpretation is used to draw descriptive conclusions in the aggregate, or motivate action at the structural level. 
Our definitions draw from literatures that use imputation to refer to disparity estimation techniques which use predicted demographic values, such as BISG~\citep{imai2022addressing,conderino2025evaluating,adjaye2014using,greengard2024improved,haas2019imputation,grundmeier2015imputing,derose2013race}; 
to our knowledge, no other body of literature consistently uses another term to describe this practice.\footnote{Other literatures~\citep{bertsimas2018predictive,ambler2007comparison} more narrowly define imputation as a semi-supervised setting where ground truth gender labels are partially available, but we do not make this specific distinction.}

We now provide some clarifications and observations along with these definitions. First, our definition of gender prediction encompasses a broad range of use cases, tasks, and contexts. It covers predicting gender for synthetic or hypothetical persons, such as AI-generated images and text, and gendering non-human objects or entities. It may also include tasks not explicitly framed as a gender prediction but used to make observations about gender. For instance, values such as unsupervised cluster membership, or a predicted proxy quantity $Z$ (e.g., hair length in images), can be treated as a substitute for gender to draw normative conclusions.\footnote{
Characterizing when a proxy can be used to draw normative conclusions is a complex question~\citep{hu2020s,hu2023race,kohler2018eddie}. 
However most antidiscrimination law is written based on prespecified identity categories~\citep{arneson2006wrongful,koppelman2006justice}. As a result, theories of algorithmic fairness and indirect discrimination are often precisely concerned with connecting proxies to identity.} This is because, e.g., substituting long hair for femininity implicitly constructs a gender prediction function.
Second, our definition of imputation can have fuzzy boundaries. A technique developed for imputation can be deployed on general prediction tasks: for example, 
BISG~\citep{elliott2009using} was developed to study and mitigate racial disparities in healthcare, but can be used for individualized ad targeting. %

Going forward, in line with our definitions, we refer to the general act of inferring gender as ``gender prediction,'' and ``gender imputation'' when specifically focusing on imputation settings, with the understanding that any illegitimacy concerns that affect prediction also affect imputation. However, the subset of prediction cases that count as imputation access a set of possible benefits (e.g., disparity measurement).\footnote{This does not preclude the possibility of other gender prediction applications outside of imputation having benefits, such as being useful to Blind or low-vision individuals~\citep{bennett2021s}, helping transgender people achieve goals around passing~\citep{chong2021exploring} or affirm them when they do~\citep{scheuerman2019computers}. These benefits also require critical evaluation, but are beyond the scope of our work.}

\section{Gender Prediction Is Illegitimate}
\label{sec:illegitimate}

In this section, we give background on gender and legitimacy, then draw from a long literature of critical scholars to explain the arguments for why gender prediction is illegitimate. This illegitimacy is serious, as it contributes to political and technological systems that enable or directly perpetrate violence, especially against transgender and nonbinary people.

\subsection{Definitions of Gender}\label{sec:what-gender}

Gender can be conceptualized in many possible ways, %
with multiple aspects (e.g., bodily attributes, gender identity, perceived gender, and gender roles)
and interactions weaved into intersectional experiences (e.g., differences between White and Black womanhood)~\citep{keyes2021you}. While broadly understood as a social construct\footnote{Social constructs are assuredly ``real''~\citep{elder2012towards}. Rather, social constructs such as gender are shaped by different societal forces; they can be constructed \textit{differently} in different contexts and changed by trends and events, as opposed to being uncontestable. \citet{haslanger2012resisting} argues that proposing and adopting definitions of social constructs can and should be driven by normative, ethical, and ultimately political, goals.}~\citep{hacking1999social}, different scholars characterize gender very differently. Debates in feminist theory about defining womanhood have spanned decades~\citep{hale1996lesbians, haslanger2012gender, jenkins2016amelioration}, with formulations as a social position~\citep{haslanger2012gender}, a performance~\citep{butler2002gender}, a ``routine accomplishment embedded in everyday interaction,''~\citep{west1987doing}, and a subjective identity~\citep{stryker2013subjugated, jenkins2016amelioration}.%

Gender identity (i.e., self-identified gender)\footnote{There are differing formal accounts of gender identity, but our work will rely on a colloquial definition as ``a sense of oneself as a man, woman, or some other gender''~\citep{jenkins2018toward}.} is one particularly important aspect of gender: 
transgender advocates argue that prioritizing other facets of gender can lead to excluding trans women from counting as women~\citep{jenkins2016amelioration}. 
However, despite both the diversity of formulations and the importance of identity, \citet{scheuerman2019computers} show that in computer vision, gender prediction and facial analysis systems de facto reduce gender down to gender presentation. Everyday understandings of gender in dominant cultures are likewise flawed, as they generally center on bodily sex and lack awareness of nonbinary genders and transgender people.

Addressing the multiplicity of gender to mitigate the harms of a single, exclusionary definition requires a degree of definitional flexibility. Just as \citet{hanna2020towards} argued for race that the use case determines the relevant meaning of race in that context (e.g., perceived race in some forms of discrimination such as racial profiling, but self-identified race in public health statistics and political mobilization), the same is true for gender.
To formulate properly trans-inclusive definitions, trans philosopher Talia Mae Bettcher~\citep{bettcher2013trans, bettcher2014trapped} 
suggests a multiple-meaning view that explicitly distinguishes between dominant (i.e., mainstream, trans-exclusionary) and resistant (i.e., trans-centered) definitions of what it means to, e.g., be a woman. While we do not adopt Bettcher's definitions exactly, we take this multiple-meaning separation of dominant and resistant definitions 
as a way to accept the primacy of gender identity while also recognizing the relevance of different aspects or definitions of gender to different settings such as discrimination, cultural norms, or structural inequality. 
In this paper, when we speak of genders such as ``man'' or ``woman'', we refer to gender identity by default. When using other formulations of gender relevant to particular ends, we refer to these as specific aspects of gender (e.g., ``perceived gender'', ``gender expression'', or ``gender as social position''); we provide a table describing these terms, the contexts to which they may be relevant, and how they are observed in App.~\ref{sec:gender-terms}.

\subsection{Definitions of Legitimacy}

Definitions of legitimacy \citep{peter2023political,beetham2013legitimation,kratochwil2006legitimacy} can be descriptive (e.g., examining if an institution, norm, or practice is accepted by a social group \citep{suchman1995managing,gerth1946politics,zelditch2001theories}) or prescriptive (e.g., normative criteria for whether an entity or practice is acceptable~\citep{stillman1974concept,adams2022concept}).
For our purposes, we focus on a prescriptive notion of legitimacy, with an emphasis on aligning with trans-inclusive gender justice. %
This necessitates diverging from majoritarian perspectives on legitimacy; as most of the world is presently cisgender%
, our framework must be capable of rejecting trans-exclusionary institutions and practices that remain widely accepted.
Instead of providing a positive definition for legitimacy, we anchor on a criterion for determining if gender prediction is \textit{not} legitimate: a gender prediction is illegitimate if it restricts individuals' agency and capacity to self-determine their gender%
. We emphasize that this notion of self-determination is not merely internal, and includes aligning external expression and social recognition to respect identity.
When a prediction is illegitimate, it should not be given governing power or social authority~\citep{peter2023political}; it has no standing to assert a person's gender, override their own account of their gender, or serve as evidence for a contradicting claim. Doing so normalizes systems that harm transgender and nonbinary people who seek to exercise self-determination.
Our definition, however, does not mean that an illegitimate prediction is never permissible to use. Rather, such a restriction on gender self-determination must be warranted by the specific purpose it serves, and only inasmuch as that purpose requires.\footnote{Consider an analogy to \citet{thomson1990realm}'s account of rights: any restriction on rights is an infringement that requires justification, but only an infringement that lacks one is a violation. We borrow this structure without adopting her specific position on rights.}
As the next section demonstrates, every gender prediction is illegitimate to some degree. In this work, we use the language of legitimacy, instead of transphobia or cissexism, to emphasize the role of this authority in gender self-determination.

\subsection{Why Gender Prediction Is Illegitimate} \label{sec:why-illegitimate}

While there are many critiques of gender prediction, we distill them into three kinds: against the act of predicting gender, against the reinforcement of gendered norms, and against the classification system of gender labels. While prior works may not originally formulate their critiques in terms of legitimacy, or these kinds, we re-interpret them here to better disentangle their validity implications, which we discuss in the next section. Our discussion centers prototypical cases of machine learning predictions of gender identity of individual humans, which we show extend to other aspects of gender and less conventional applications.

\subsubsection{Gender Predictions} %
The process of assigning a gender label is illegitimate, especially for individuals.
First, automatically assigning and policing a person's gender further entrenches administrative harms: \citet{katyal2021gender} link gender, the state, and surveillance capitalism, where predicting gender further empowers government and corporate institutional control (e.g., in suppressing LGBTQ+ content on social media). \citet{costanza2020design} names the privacy-violating experience of being searched by the TSA as a trans woman, and \citet{keyes2018misgendering} warns of the possibility of gender prediction being used in bathrooms, with potentially violent enforcement. Of course, not all applications of gender prediction actively enable violence, but all applications must assess these kinds of risks.

Second, the actual process of prediction is harmful. Misgendering, or erroneously assigning a gender to a person, directly harms~\citep{kukla2023telling, howansky2022him, kapusta2016misgendering,mclemore2018minority,matsuno2024default} individuals, especially (but not limited to) transgender people. As a second-order effect, machine learning exacerbates this problem: the seeming objectivity of technology normalizes gender predictions by humans, and can enable intentional transphobia. 
Note that misgendering is harmful even when referring to aspects of gender such as gender expression; telling someone that they do not look like their gender identity is generally still a dignitary insult. Less rigid categories or labels may lead to less severe misgendering acts (e.g., describing someone's appearance as ``more masculine'' is a less definitive claim), but remain inherently presumptive.

Third, even if a prediction is accurate and does not misgender, it hinges on the assumption that gender is something that can be predicted. This presupposes a definition of gender that decenters gender identity, conflicting with trans-centered principles of gender self-determination. This
disrespects the inherent, first-person authority people have over their gender~\citep{bettcher2009trans}. Similarly, \citet{keyes2018misgendering} categorically rejects the idea that gender prediction is compatible with trans identity, writing, 
``a trans-inclusive system for non-consensually defining someone’s gender is a contradiction in terms.'' Even when claims are explicitly scoped but about other aspects of gender (e.g., perceived gender), this framing can still privilege externally assigned perceptions over individuals’ own gender identities.

\subsubsection{Gender Norms} %

Gender prediction reinforces gendered norms and associations between observable features and a person's gender.
Technologies using biometric data, such as facial analysis, are built on especially harmful bioessentialist norms (i.e., they equate gender and sex)~\citep{schumann2021step}. While other norms such as gendered names are less egregious, gendered language associations can still cause harm~\citep{gautam2024stop}. These associations make it difficult for a person to have their gender identity respected, or align others' perceptions of their gender with their identity. Moreover, associations can vary between cultural context and race: infamously, \citet{buolamwini2018gender} found that gender prediction using facial analysis is less accurate for Black women.

Moreover, while humans can also predict gender via harmful gender norms, using technology introduces new harms. Algorithmic gender prediction changes the fundamental dynamics of gender performativity, which makes gendered norms more difficult to contest and harder to change.
As \citet{butler2002gender} describes, gender is a feedback loop of performance. Every act of gender is performed in response to the expectations that gender itself creates, projecting norms from the past onto the present. However, while human interaction is full of opportunities to shift and contest these gendered expectations, algorithmic systems are fixed by default, and model updates occur at the discretion of the developer. 
To give an example, \citet{chong2021exploring} describes cases of transgender people using automated gender prediction to see if they pass as their gender (i.e., appear as their preferred gender presentation). These attempts are tied to a fixed decision boundary between man and woman. Transgender individuals must force themselves to fit an algorithmic standard to be perceived as a particular gender, instead of being empowered to change such standards.

\subsubsection{Gender Labels} %

Gender prediction reinforces gender as a restrictive system of classification. Whether in bureaucracy or in databases, many gender schemas are insufficiently expressive, and either erase genders other than ``man'' and ``woman,'' or homogenize them (i.e., grouping genderqueer, nonbinary, and gender non-conforming together). Those who do not fit into these classifications end up in residual categories that silence them~\citep{star2007enacting,bowker2000sorting}. Trans men and women, too, face barriers to changing their categorization in this system, even if the categories of ``man'' and ``woman'' exist.

Using these classifications in systems of governance harms vulnerable individuals. Critical work by transgender advocate \citet{spade2015normal} coins \textit{administrative violence} to demonstrate how the administrative state 
enables and enacts violence on transgender and nonbinary people. \citet{hoffmann2021terms} extends this analysis to technological systems with \textit{data violence}, as data-driven systems reinforce the discursive structures that normalize violence against those on the margins. 
Note that labels not based on identity, or not applied to real and whole individual people, still reinforce classification systems that limit gender self-determination (e.g., the arbitrariness of attempting to censor ``female-presenting nipples''~\citep{pilipets2022nipples}).

\section{Gender Imputation May Yield Valid Disparity Measurements}
\label{sec:valid}

In the previous section, we provided an overview of reasons previous scholars have argued for the illegitimacy of gender prediction, and thus the harms it causes. We wholly agree with those arguments, but here argue that in the  \textit{imputation} case of predictions, the illegitimacy of the imputation does not necessarily indict the validity of measurements based on those imputations. Moreover, we argue that some forms of discrimination are inherently built on different aspects of gender (e.g., those against the perceived gender of individuals), and can by definition be imputed and inferred.
Certainly, this does not mean that \textit{all} gender imputation is valid, and many existing applications fall critically short: we provide recommendations in these cases.

\subsection{Measurement Validity}

Validity allows us to determine what justified claims can be made from a particular measurement~\citep{cronbach1955construct,jacobs2021measurement}. \citet{adcock2001measurement}'s framework distinguishes a \textit{background
concept}, encapsulating different notions of disparity (e.g., unequal treatment, unequal opportunity, unequal representation) across different meanings of gender (including those in Sec.~\ref{sec:what-gender}), from a \textit{systematized concept}, the single specific formulation or definition of gender disparity that is the focus of a measurement (e.g., ``the wage gap in feminized occupations in the United States''). \citet{wallach2025position} apply this framework to the evaluation of generative AI models and draw on \citet{messick1993foundations} in suggesting various criteria of validity be applied to examine a measurement. For brevity, we touch on three criteria here that are especially salient.
Content validity requires that measurements draw upon the appropriate observed properties and dimensions that make up the concept. Convergent validity examines if a measurement correlates with other established validated measurements. This provides an empirical way of validating disparity measurements that use imputed gender: comparing them to other measures of discrimination, such as in settings where self-reported gender is available. Lastly, consequential validity, which relates to our conception of legitimacy, requires that we weigh the consequences of how a measurement shapes the quantities we study, the conclusions we draw, and ultimately, the world.

Of course, making any judgments weighing harms and benefits requires showing that disparity measurements using imputed gender can be valid enough to achieve benefits, which requires careful consideration and design. As \citet{wallach2025position} note, machine learning researchers and practitioners rarely provide sufficiently precise specifications for the systematized concept of interest. 
Naïve approaches to gender imputation, via conventional prediction models, will almost necessarily be invalid.
To understand the validity of gender disparity measurements based on gender prediction, we consider the conceptualization and operationalization steps of the measurement process.

\subsection{Conceptualizing Gender for Measuring Discrimination} \label{sec:conceptual}

Conceptualizing (i.e., ``systematizing,'' in \citet{adcock2001measurement, wallach2025position}'s terms) gender in the fairness context requires intentionally narrowing its scope, as 
fairness and gender are both essentially contested concepts~\citep{gallie1955essentially} that require
narrowly scoped \textit{pragmatic measurements}~\citep{hand2004measurement}.
As we have shown in Sec.~\ref{sec:what-gender}, gender has many different formulations. Similarly, there are many definitions and disagreements about what constitutes discrimination~\citep{hellman2008discrimination,kohler2018eddie,alexander1992makes}.
Just as \citet{hanna2020towards} observe with race, we take the position that an effective conceptualization must consider what aspects of fairness are relevant to gender and what aspects of gender are relevant to fairness.
Both are dependent on the input data available and understandings of the context. 

\subsubsection{Conceptualizing Fairness} %

Debates about algorithmic fairness measures arise from differences in substantive value judgments about fairness and justice. 
No one measurement or concept can capture \textit{all} notions of fairness~\citep{blodgett2020language}; \citet{jacobs2021measurement} suggest that we need only be precise about which aspects or purposes motivate us and recognize when they are appropriate. For our purposes, we chiefly focus on disparity as a basic quantitative measure frequently used in the algorithmic fairness and audit literature.

Gender discrimination has multiple dimensions. Trans author Julia Serano distinguishes between two kinds of discrimination central to her experiences as a trans woman: 
she coins \textit{traditional sexism} to refer to sexism that privileges masculinity over femininity and targets those who are female or feminine, and \textit{oppositional sexism} for sexism that includes transphobia, homophobia, and cissexism and targets gender deviance~\citep{serano2007whipping}.
Further, these two kinds of discrimination are not mutually exclusive, and can compound into what she terms \textit{trans-misogyny}; as Serano writes, %
``I am not dismissed for merely failing to live up to binary gender norms, but for expressing my own femaleness and femininity.''%
Traditional sexism is a core component of the discrimination affecting some of the most marginalized members of the trans and nonbinary community.

Those concerned with oppositional sexism may be critical, and justifiably so, of those concerned with traditional sexism who employ gender imputation, since that practice will invariably fail to respect transgender and nonbinary identity, likely fail to detect discrimination against nonbinary people, and thus implicitly privilege one form of discrimination over the other. Imputation always has oppositional sexist harms, and we agree that gender imputation, even when valid, should face heavy scrutiny.
However, %
allowing for the \textit{possibility} of imputation yielding valid measurements, without claiming it is legitimate, makes addressing traditional sexism, including instances of trans-misogynistic discrimination, more tractable.%
Thus, while we primarily discuss imputation's suitability for measuring traditional sexism, we hope that disentangling traditional from oppositional sexism also highlights how the latter has been obscured by generic accounts of sexism. %
We proceed with the understanding that both forms of sexism are important, and that the illegitimacy of imputation is a reason that researchers should develop different measurements and techniques to study oppositional sexism.

\subsubsection{Conceptualizing Gender} %

Note that the validity of a disparity measurement using imputed gender differs from the validity of \textit{imputed gender as a measurement itself}, and the validity of one does not inherently determine the other. However, they are in many cases intertwined: for example, measuring discrimination that causes representational harms may straightforwardly involve counting the number of women, or people perceived as women, in a photo.

For this reason, we describe how to choose an appropriate conceptualization of gender that is relevant to a particular measurement of discrimination. 
Drawing from \citet{hanna2020towards}, the conception of gender relevant for discrimination is often different depending on the context. For instance, for hiring discrimination in an interview, visually perceived gender is relevant. For hiring discrimination in resume screening, perceived gender from name is relevant. In both cases, self-identified gender identity is less relevant, though of course it bears on presentation through performance.

By conceptualizing gender under these more measurable formulations (e.g., perceived gender, gendered social position) compared to gender identity, which cannot be predicted, we are better suited to make less illegitimate claims and make more valid measurements. 
Of course, these more measurable formulations are distinct from and can conflict with gender identity, and such conflict is itself a harm for the reasons listed in Sec.~\ref{sec:why-illegitimate}. 
Having a feminine perceived gender does not make someone a woman. Distinguishing these formulations, and what an imputed label implies, minimizes the illegitimacy of making a prediction and specifies a more validly measurable concept.

In fact, using these other conceptualizations of gender not defined by self-identification can, at times, even better align with our conceptualization of discrimination and lead to greater validity. 
For example, misogynistic street harassment generally occurs without regard for a person's gender identity; a nonbinary person misperceived as a woman might still experience catcalling. In such a case,
being perceived\footnote{This raises the question: perceived by whom? Often, the perception most relevant for discrimination measures is that of a possibly discriminatory decision-maker. Of course, in most contemporary cases and at scale, discrimination is more complex. Nonetheless, part of what makes structural discrimination what it is is that it is widespread by definition. 
} or socially situated as a woman is sufficient to be subject to traditional sexism.
Moreover, consider that people perceived as not conforming to gender norms can experience harms that transgender and nonbinary people perceived as conforming do not, regardless of whether they are cisgender~\citep{mary1995disaggregating}. %
Thus, 
imputation that ``recogniz[es] performative markers of underrepresented genders'' could measure some forms of oppositionally sexist discrimination~\citep{scheuerman2019computers}.
For this reason, we conclude that disparity measurements using gender imputation for conceptualizations of gender other than gender identity can validly systematize traditional sexism, even if imputed values do not reflect individuals' gender in the broad sense.

\subsection{Distinguishing Legitimacy from Validity} \label{sec:distinguishing}

We clarify here that legitimacy and validity are distinct qualities. Consider a hypothetical algorithm used by an oppressive state that successfully classifies certain individuals into some category in order to treat them poorly: its validity may not be in question, but its authority is illegitimate regardless. We provide a brief two-by-two table distinguishing cases of valid and invalid, and legitimate and illegitimate, gender disparity measurements in App.~\ref{sec:leg-val}.

In examining the validity of disparity measurements that use imputed gender, we make the case that gender imputation is inherently illegitimate, but not in all cases invalid. Of course, gender imputation is \textit{often} invalid for the same reasons that it is illegitimate. For example, classification systems of gender, alongside being harmful and violent, are simply \textit{inaccurate}; gender norms are often far less stable and inherent than model designers assume; and nonbinary people cannot be accurately classified~\citep{scheuerman2019computers,keyes2018misgendering}. 

However, there are cases where a gender imputation may hold some validity. For example, we believe that the experience of feeling like a gender minority in computer science spaces is justified, and grounded in reality, even without knowing others' gender identities. These feelings are grounded in aspects of gender such as perception and social position that are by definition constructed by inferences; rejecting these feelings inherently dismisses these experiences.
Of course, such an experience does not justify misgendering others or void harms from doing so. To reconcile these conflicting intuitions, we distinguish illegitimately assuming others' genders from the feeling of ``not seeing people who look like me,''
and distinguish their relevance to the oppositional sexist harm of misgendering from the traditional sexist harm of a misogynistic culture. 

Furthermore, taking the position that gender imputation is always invalid may cause or enable even greater harm. 
\citet{boyd2022differential} point to cases where the correctness of statistical procedure becomes a political weapon --- often by those seeking to harm transgender and nonbinary populations. For example, the CFPB's use of imputed race~\citep{bureau2014using} was accused of being statistically invalid~\citep{baines2014fair} by corporations and political actors who sought to curtail 
their anti-discriminatory regulatory powers and weaken the legal foundations of anti-discrimination law%
. 
Similarly, the harms of undetected discrimination may trade off against the harms of using gender imputation when the former is difficult or impossible to quantify by other means. Such judgments can only be made if we allow for the possibility of assessing the validity and therefore effectiveness of imputation. In the next section, we provide recommendations for assessing and improving this effectiveness, and determining when deploying imputation might be warranted in spite of its inherent illegitimacy.

\section{Recommendations}
\label{sec:recommendations}

We have argued that gender imputation is always illegitimate, and therefore harmful and unsuitable for measuring oppositional sexism, but may sometimes be valid for measuring discrimination related to traditional sexism. %
However, while we strongly argue for acknowledging the validity of gender disparity measurements based on imputation, weighing if or when an application of gender imputation is justified is a separate question. Making this judgment requires exploring whether alternative ways of characterizing gender fairness exist, as well as the possibilities of minimizing illegitimacy while maximizing validity and benefit, where the status quo is currently insufficient in both regards. Here, we provide recommendations toward these ends when constructing measurements that use imputed gender.

\subsection{Weighing Tradeoffs between Illegitimacy and Validity} 
\label{sec:rec-consequence}

We draw on the principle of beneficence from the 1979 Belmont Report~\citep{belmont1979}, which established ethical guidelines for research, to suggest that we should consider how to 1) maximize possible benefits and 2) minimize possible harms. 

A justified application of gender imputation should bring otherwise unachievable benefits. Note that the validity of a disparity measurement is instrumental to achieving benefits, but not necessarily a benefit in itself. 
Even validly measuring disparities and showing them to be normatively wrong may not constitute a substantive benefit if such a diagnosis does not lead to downstream awareness, actions, or impacts. An imputation model that is carefully defined and perfectly accurate remains unacceptable if the audited task is normatively trivial.
We suggest weighing harms against the consequences of inaction: leaving gender unexamined in settings without self-reported data or other mechanisms for detecting discrimination may also risk severe harm. 

Harm minimization, on the other hand, can help to address some legitimacy issues.
As imputation will always be to some degree illegitimate, any usage must minimally restrict gender self-determination.
A general-use imputation of gender as a whole cannot be justified over using a more narrowly tailored gender concept, and imputation cannot be justified if there are reasonable alternatives for calculating disparity or acquiring gender data (e.g., joining records with self-identified sources~\citep{bennett2021s}).
Minimizing any delegitimizing factors also maximizes consequential validity.
For instance, making predictions on vulnerable populations and enabling downstream misuse (e.g., by justifying new surveillance infrastructure) 
are highly illegitimate and harmful uses, as discussed in Sec.~\ref{sec:illegitimate}. We suggest the following application-level recommendations in the rest of this section to maximize benefit and minimize harm when judging if imputing gender to measure disparity is worthwhile.

\subsection{Measurements Should Use a Narrowly Scoped Conceptualization of Gender}
\label{sec:rec-scope}
Examining a specific task, such as detecting gendered algorithmic bias in resumes, requires operationalizing a contextual and narrow conceptualization of gender. An imputation technique here should not predict a general gendered social position or perceived gender as a static property of a person, but rather that person's gendered social position or perceived gender in that particular setting and set of norms. For example, \citet{haslanger2012gender} argues that a man of color may not be socially positioned as a man when experiencing emasculating racial violence. Similarly, perceived gender depends on name, appearance, voice, occupation, social role, and varies with which properties are observed; for instance, a cisgender woman named ``Alex'' might be assumed to be a man from her name, but a woman from her appearance. 

Strengthening content validity requires accounting for the complexity of these gendered norms and relationships. Moreover, these gendered positions or perceptions that features imply also vary between time, context, and culture. For example, the early days of computing were led by women~\citep{light1999computers,abbate2012recoding}, and marriage comes with different socially positioned consequences in the modern day than in historical settings. Perceptions can also vary; a dataset of names drawn from Spanish speakers will have different gendered associations with the masculine name ``Juan'' than an English-language dataset of mainland Chinese names, where a common feminine character is romanized the same way. 

Imputation techniques must also actually align with this narrow conceptualization by using appropriately defined and collected data and minimizing the scope of claims made about an individual. For instance, combining two name datasets when one is associated with sex assigned at birth and the other with self-reported gender~\citep{mohammad2020gender} illegitimately conflates sex and gender and invalidates any downstream disparity measurement. Instead, imputation models should be trained and validated by collecting perceived gender data from names or resumes directly, such as by surveying recruiters.
Properly collected data also allows for a consistent interpretation of what an imputation \textit{is}. If an imputation only predicts ``perceived gender based on a resume,'' for example, this treats gender less as a universal constant property of a person, and only says something about an aspect of their gender in a particular context. While still illegitimate, we consider this less illegitimate than attempting to impute a person's gender as a whole.

\subsection{Imputations Should Use Appropriate Inputs}
\label{sec:rec-features}
Given a properly scoped conceptualization of gender in a particular setting, any imputed values and downstream disparity measurements should use appropriate and valid features that reinforce less illegitimate norms.
For example, in social media data, \citet{scheuerman2019computers} observe that simple image-based gender classifiers ignore captions or visual artifacts that state or imply someone's gender, which both disrespects users' agency and ignores salient information. In other cases, methods may rely on irrelevant features. \citet{meister2023gender} find that spurious visual artifacts, such as particular background colors, prove predictive of gender even when the person is entirely removed from the image.
Both using inappropriate features and ignoring appropriate features are concerns in facial analysis imputation methods. Using biometric data is not only illegitimate for conflating sex with gender, but also invalid, as people often perceive gender through other aspects of appearance. While not without its own issues, incorporating performative aspects of gender presentation that individuals have more agency over, such as clothing, makeup, or hairstyle would improve both validity and legitimacy~\citep{scheuerman2019computers}. 

\subsection{Gender Imputations Should Be Aggregated} \label{sec:rec-aggregation}

Disparity measurements that use imputed gender should be examined in the aggregate to avoid drawing conclusions about individuals, and calculated with statistically sound methods.

Using imputed gender to calculate aggregate measurements of disparity (e.g., the difference in hiring conditional on gender) can be less illegitimate than using imputed gender values for individual-level decisions. When showing a gender disparity with an aggregate measurement, we minimize reliance on, and authority given to, individual predictions. 
Intuitively, a social environment such as a computer science conference can broadly feel dominated by men, or devoid of people with a similar gender presentation. This is a discriminatory phenomenon that exists only in the aggregate. While gender is nonetheless perceived, this discriminatory experience is built on a holistic set of observations without hinging on a strong claim about any individual's gender.
Aggregate measures are more easily understood as imperfect estimates and leave room for individual-level uncertainty, contestation, and variation.

Moreover, disparity measurements that properly aggregate imputed values can maximize empirically assessable forms of validity. 
Performing well on these forms of validity requires using best practices in calculating disparity measurements.
\citet{chen2019fairness} and \citet{dong2025addressing}
observe that using raw, continuous model outputs, instead of discretizing predictions into the distinct categories used in self-reported data, as is commonly done, can mitigate statistical biases in aggregate disparity measures.\footnote{Note that these statistical properties rest on a model calibration assumption. Calibration is often used to interpret model outputs as probabilities, but we do not necessarily endorse this interpretation as either philosophically rigorous \citep{hu2025does} or trans-inclusive. Of course, any mitigation of estimation bias, like calibration itself, can be empirically tested in a given context by gathering real data.} Empirically, these methods lead to results that more closely match results from self-reported data, showing better convergent validity.

\section{Case Studies}
\label{sec:case-studies}

Lastly, we examine three gender imputation tasks as case studies, differing in their target population and feature modality.

\subsection{Auditing Stereotypes in Image Generation} \label{sec:case-generation}

\textit{Background.}
Generative text-to-image models are known to output images that are not representative nor diverse~\citep{wan2024survey}. 
Most commonly for bias evaluations on these models, researchers examine correlations between gender and occupation: generated images of doctors or managers disproportionately look like men, while images of nurses and secretaries look like women~\citep{luccioni2023stable, naik2023social}. This brings out our conflicting intuitions: explicitly and formally recognizing this bias in the image generation setting requires being able to articulate what it means to ``look like men'' --- thus motivating gender imputation.

\textit{Legitimacy.}
Since generative models create images that do not reflect real individuals,
not all the harms of making a gender prediction apply.
Gender prediction models cannot misgender generated images of people, because there is 
no real person to experience the dignitary harms of misgendering or have authority over a gender identity. Of course, predictions still reify gender norms, labels, and normalize the act of prediction, so many other concerns in Sec.~\ref{sec:why-illegitimate} apply. Using methods that less directly perform gender classification can also help avoid normalizing these practices. For example, work that builds a reference set of image clusters for different genders, then assigns generated images to those clusters based on embedding distance, sidesteps some implications of direct classification~\citep{luccioni2023stable}. Still, completely avoiding interpretations of prediction is likely impossible.

\textit{Validity.}
The lack of real individuals in generated images means that only aspects of gender such as perceived gender exist. This aligns with our earlier conceptual distinctions: gender identity is not validly measurable via imputation, but other aspects of gender may be.

\subsection{Auditing Gender Disparities in Film} \label{sec:case-film}

\textit{Background.}
Another setting examines gender representation in film~\citep{jang2019quantification,schmidt2021exploring}, by predicting gender frame by frame to identify disparities such as, e.g., differences in total screen time between men and women, often using computer vision.

\textit{Legitimacy.}
Computer vision prediction of gender has many illegitimate dimensions, with few factors that can mitigate this illegitimacy.
The established arguments in Sec.~\ref{sec:why-illegitimate} apply here, although some factors are worth particular notice. Unlike the case of generated images, concerns about misgendering harms apply in full force. As a particularly notable harm, two works that perform this analysis chiefly use face-based gender prediction models, which rely on particularly illegitimate gender norms~\citep{jang2019quantification, schmidt2021exploring}. 
And, while \citet{schmidt2021exploring} includes an ``androgynous'' classification label, this may not substantively address issues with gender labels; their ``androgynous'' label confusingly refers to ``either  multiple genders on one screen or uncertainty by the model,'' with no substantive discussion of gender nonconformity or the implications of ambiguity. %

\textit{Validity.}
The validity of the relevant disparity measures here is closely tied to the validity of the imputation method itself, which may be more limited because of the previously discussed issues with facial prediction. While visually perceived gender is relevant and valuable for this analysis, audiences generally see far more than only a face, and perceive gender from a much more expansive set of cues (e.g., the sound of a voice, the gender of address used in dialogue) than still frames. Arguably, with consideration to the ethical concerns of facial recognition technologies~\citep{stark2019facial}, directly using facial recognition to identify specific actors and matching to their biographies or character descriptions would likely provide more valid measurements.

\subsection{Imputing Gender from Names}  \label{sec:case-names}

\textit{Background.}
Another common gender prediction task is inferring gender from names, as given names in many languages are gendered. For example, \citet{vogel2012he} and \citet{mohammad2020gender} use various sources of name-gender associations (e.g., the US Social Security Administration's list of baby names, US Census publishing, manual data collection) to examine academic authorship and citation. Recent approaches use machine learning for this prediction task, often using social media data~\citep{santamaria2018comparison,hu2021s,tang2011s,sebo2021performance,karimi2016inferring}. We also highlight the resume audit experiment design introduced by \citet{bertrand2004emily}, who manipulate names on resumes and submit them to real-world job postings. Drawing conclusions about perceived gender, of course, requires gendering names.
\citet{gautam2024stop} analyze the practice of name-based sociodemographic prediction in detail, discussing concerns in both ethics and validity.

\textit{Legitimacy.}
\citet{gautam2024stop} %
suggest a gradation of ethical permissibility depending on the application. Notably, they find gender imputation on real people unacceptable.
Nothing substantively mitigates the illegitimacy of this practice --- name-based predictions are uncontestable, normalizing, and misgendering, and often use associations based on sex.
In addition, \citet{gautam2024stop} address predictions not tied to real people --- that is, to draw conclusions about gender bias ``based on how NLP models handle synthetic names of people assumed to be exclusively female.'' Similar to the generative model case in Sec.~\ref{sec:case-generation}, such predictions are marginally acceptable as they do not risk misgendering, but remain concerning as they ``entrench hegemonic folk theories of names and people’s identities.'' %

\textit{Validity.}
\citet{gautam2024stop} identify four barriers to validity: %
1) biases arise from mapping names exactly onto gender; 2) researchers conflate different aspects of gender; 3) evaluating gender prediction systems requires self-reported data, with poor generalizability; and 4) echoing consequential validity concerns, the construction of a classification system shapes its conclusions. 
We consider whether these issues could in theory be minimized. First, as discussed in Sec.~\ref{sec:rec-aggregation}, 
using aggregated measures with continuous estimators can mitigate discretization-based biases.
The second and third issues can both be minimized by aligning imputation practices with a clear conceptualization of gender, as discussed in Sec.~\ref{sec:rec-scope}. Indeed, mapping names to sex assigned at birth conflates different gendered concepts; however, \citet{gautam2024stop} concede that validity is stronger when examining perceived gender, which we further qualify to \textit{perceived gender based on written name}. Scoping the imputed gender concept to ``perceived gender based on name'' can also warrant recruiting %
annotators from a fitting background (e.g., job recruiters in anglophone companies) for more robust error evaluation. Lastly, the consequential concerns with classification warrant the need for clarity in conceptualizing discrimination. However, this issue is inherent to \textit{all} studies of discrimination; working within a flawed system does not inherently invalidate findings of disparity, and must be weighed against not identifying discrimination at all (Sec.~\ref{sec:rec-consequence}). 

\section{Discussion}
\label{sec:discussion}

Our work has argued that gender imputation is illegitimate but may be used for valid disparity measurements which pertain to traditional sexism. While considering any particular instance requires care and nuance, we provide ways to weigh and evaluate any benefits imputation may bring against its inherently harmful illegitimacy. Our analysis critiques existing imputation practices and resolves conceptual tensions in the normative obligations around gender prediction.

\subsection{Takeaways}

Thematically, we emphasize two takeaways. First, being precise about gender imputation highlights the pervasiveness of trans-exclusionary assumptions. %
Taking a moral prohibition against gender prediction seriously requires addressing cases where well-intentioned or epistemically valid motivations exist.
Second, there are forms of discrimination that are based on other aspects of gender. While self-identification and gender identity deserve primacy, they do not capture \textit{all} facets of how gender manifests in the world or in individual experience. Consider that many trans people experience misogyny without being women.
Engaging with the nuances allows us to identify discrimination (e.g., on the basis of a feminine appearance) without \textit{ascribing} an identity (e.g., on the basis of being a woman). 

More practically, our work has implications both for advocates and for algorithmic fairness practitioners.
Although many gender prediction use-cases are poorly motivated, harmful, and irredeemably flawed, clarity about why the practice is illegitimate can help set the bounds of what is permissible in gender bias research (such as the guidelines provided by \citet{scheuerman2019computers} and \citet{gautam2024stop}). 
Furthermore, without precise and careful engagement with trans philosophy and other socially just ends, concerns about misgendering and oppositionally sexist harms will continue to be sidelined as unreasonable or easily outweighed by other concerns. We observe that this lack of clarity has led to practitioners acknowledging these harms in theory but failing to navigate them in practice. Our contributions here provide tools to navigate this tension, highlighting the need for methods to capture oppositional sexism, beyond simply rejecting imputation.

For researchers or practitioners who may use imputation,
our recommendations argue that existing practices should be held to a higher standard. 
By showing that more valid, and less illegitimate, measurements are possible, our work shows that merely making perfunctory acknowledgments that gender prediction is limited and trans-exclusionary is insufficient~\citep{devinney2022theories}. Many unnecessarily illegitimate practices cannot be justified merely because bias assessment provides some benefit. While our recommendations may require more effort, building fair systems takes both imagination and commitment.

\subsection{Limitations}

While we provide recommendations for improving validity and reducing illegitimacy, we recognize this list is not exhaustive. There is more work to be done in creating imputation tools that can validly measure discrimination, and reduce harms stemming from their illegitimacy. Future work may also broaden the scope of our initial argument here: we focus on a relatively narrow kind of discrimination (i.e., disparity), for a few aspects of gender. 
Critically, we echo prior work by \citet{scheuerman2019computers} calling for developing techniques that can measure oppositional sexism. %
In clarifying the bind of gender imputation, we hope it brings to light the neglect of evaluations for oppositional sexism. More thoroughly drawing connections to and conclusions about other demographic attributes is also beyond our work. Engaging with more complex theories about how definitions of discrimination and gender are intertwined or more thoroughly challenging the need for demographic data in algorithmic fairness are other normative assumptions made here that can and should be questioned.

\section{Conclusion}

In this article, we identify and work through conflicting intuitions around algorithmic gender imputation by decomposing them into questions of legitimacy and validity. This separates a normative prohibition of gender prediction due to its illegitimacy from an epistemic incongruity with a trans-inclusive understanding of gender. Our analysis shows that gender imputation is illegitimate, like all gender prediction, but can be useful in describing real forms of discrimination. To reconcile the normative implications around deploying imputation, we provide the means to analyze and minimize the harms inherent to gender imputation while maximizing the potential benefits from a particular context. We recommend practices to navigate this tension by minimizing the harms of illegitimacy and achieving benefits from maximizing validity. Our work charts a path toward identifying discrimination without reifying the harmful assumptions that underlie it. With this, we can better address the core issues in traditional sexism, while also casting light on the overlooked needs of oppositional sexism.

\section*{Positionality Statement}
Both authors of this work are queer researchers embedded in the algorithmic fairness research community. The first author is a genderfluid nonbinary person affected by trans-misogyny, and their perspective in writing this paper is informed by their personal experiences with having their gender presentation perceived differently in different contexts, and the limitations of articulating that experience in sociotechnical accounts of data collection and algorithmic discrimination.

\section*{Ethical Considerations Statement}

Our work discusses transphobia and trans-exclusionary ideas on technical and philosophical levels, where uncritically repeating harmful lines of reasoning can normalize transphobia. We considered multiple formulations of our argument and ultimately chose this particular framing as one most likely to achieve our goal (addressing the bind we describe in the paper) while minimizing the chance of justifying harmful practices or ideas, which we describe more in the adverse impacts section.

\section*{Adverse Impacts Statement}
As discussed in our ethical considerations statement, we were concerned with three main adverse impact risks: 1) our work being twisted to justify harmful uses of imputation, 2) contributing to a discourse that decenters trans-inclusive definitions and aspects of gender, and 3) enabling harmful research practices even when an application is justified (i.e., insufficiently minimizing harm). To mitigate the first, we formulated the argument to be as narrowly tailored as possible: we avoid contesting the ethical injunction against gender prediction by distinguishing legitimacy from validity, we narrowly scope the definition of imputation, and we formulate our arguments to only be relevant to discrimination to minimize how easily they could validate other applications of imputation, alongside explicitly saying in the paper that most imputation cases will still be unjustified. The second we address by directly stating our normative commitments to trans inclusion in our discussion of gender and shaping our argument around them. For the third, we attempted to balance both provocation and practicality with our recommendations. We judged it valuable to provide a mix of recommendations that may be more immediately implemented and those that might require greater resources, so as to not enable practitioners to stop at ``good enough;'' for this reason, we also do not treat these recommendations as exhaustive.

\section*{Generative AI Usage Statement}

We used Claude Sonnet 4.5 and Claude Opus 4.5 in our literature review to broaden our search for relevant works and format the table in the appendix. We also used Claude Opus 5 for feedback on grammar, wording, and argument structure.

\section*{Acknowledgments}
We thank Aniket Kriplani, Aurora Zhang, Gena Kim, Meera Desai, Nikhil Garg, Thalia Zhang, Qiaoying Chen, Rui-Jie Yew, Vyoma Raman, members of the Cornell AI Policy and Practice Initiative, participants of the 2025 AI and Human Values workshop organized by the Central NY Humanities Corridor, and several anonymous reviewers for their feedback in developing this paper. We are particularly grateful to Daniel Susser for his persistent encouragement and support.

ED acknowledges that this material is based upon work supported by the National Science Foundation Graduate Research Fellowship Program under Grant No. DGE-2139899. Any opinions, findings, and conclusions or recommendations expressed in this material are those of the author(s) and do not necessarily reflect the views of the National Science Foundation.
AW acknowledges support from Mastercard and the Survival and Flourishing Fund.

\FloatBarrier
\bibliography{bib}
\newpage
\appendix

\section{Table of Gender Terms} \label{sec:gender-terms}

In this section, we provide Table~\ref{tab:gender-terms}, which overviews many of the aspects of gender we describe, as well as other terminology useful for drawing basic distinctions in gender and sex.

\begin{table*}[tb]
    \centering
    \begin{tabular}{p{6cm}p{5cm}p{5cm}}
    \hline
        \textbf{Gender Term} & \textbf{Description} & \textbf{Relevant Contexts and Outcomes} \\
    \hline
        Gender Identity & Subjective, self-defined, often anchored to gender categories & Individual experiences, gender labels, and gender as a concept, anti-trans discrimination \\ \\
        Gender Norm & Associations between observable characteristics (e.g., appearance, social position) and gender categories & Structural patterns, social interactions, stereotyping \\ \\
        Femininity / Masculinity & Characteristics associated with the gender categories ``woman'' / ``man'' via gender norms & Stereotyping, social expectations, characterizing gender labels \\ \\       
        Gender Expression & Observable characteristics person exhibits, driven by gender identity and gender norms & Behavioral patterns \\ \\
        Gender Presentation & Observable characteristics a person exhibits in order to be perceived as a particular gender, such as clothing, cosmetics, and tone of voice & Passing, interpersonal interactions, self-image \\ \\
        Perceived Gender & What others observe of a person's gender, broadly construed & Interpersonal interaction, discrimination \\ \\
        \quad - Visually Perceived Gender & What others observe of a person's gender based on appearance & Representational harms in visual media, sexual harassment \\ \\
        \quad - Perceived Gender Label & A gender label associated with a perceived gender & Interpersonal ascriptions, misgendering and social categorization, discrimination \\ \\
        \quad - Perceived Femininity / Masculinity & How observations in perceived gender relate to femininity / masculinity & Stereotyping, social expectations, discrimination \\ \\ 
        Sex & Bodily characteristics such as genitalia, hormones, and genetics & Healthcare disparities, reproductive justice \\ \\
        Gender as a Social Position~\citep{haslanger2012gender} & ``How one is viewed, how one is treated, and how one's life is structured socially, legally, and economically,'' focusing on power and hierarchy & Political economy and political advocacy, labor, state governance, the carceral system \\ \\        
    \hline
    \end{tabular}
    \caption{We summarize the language we use to describe several aspects of gender and sex, broadly construed, drawing heavily from \citet{hanna2020towards}'s table of race. We note that this list is not exhaustive, and other authors may use different terms,  descriptions or definitions (e.g., \citet{jenkins2018toward} offers a rigorous account of gender identity in much greater detail). 
    }
    \label{tab:gender-terms}
\end{table*}

\section{Table of LGBTQIA+ Terms and Language} \label{sec:queer-terms}

In this section, we provide Table~\ref{tab:queer-terms}, which collates LGBTQIA+ terms used in our paper, as defined by a crowdsourced wiki run by LGBTQIA+ community members.

\begin{table*}[tb]
    \centering
    \begin{tabular}{p{8cm}p{8cm}}
    \hline
        \textbf{Term} & \textbf{Definition} \\
    \hline
        \textbf{cissexism} & A form of sexism, specifically, a way of thought in which only cisgender people are seen as normal or right. Cissexism is harmful to all kinds of transgender people, including nonbinary people. \\ \\
        \textbf{misgender} & To address someone in a way that contradicts their gender identity. This can be accidental, but if intentional, it can be an example of discrimination against transgender people (cissexism). \\ \\
        \textbf{non-binary gender, nonbinary gender, or nonbinary} & An umbrella term for all who don't identify as just female or male. Though there are many kinds of nonbinary gender identities, some people identify as ``nonbinary'' only. \\ \\
        \textbf{trans} & Short for transgender or transsexual. \\ \\
        \textbf{transgender} & An umbrella term for those with gender identities that don't match the genders they were assigned at birth. \\ \\
        \textbf{trans-misogyny} & Discrimination and hate crimes against transgender women. \\ \\
    \hline
    \end{tabular}
    \caption{All terms are directly sourced from the community-run Nonbinary wiki's English glossary of gender terms.
    }
    \label{tab:queer-terms}
\end{table*}

\section{Table of Legitimacy and Validity Examples}  \label{sec:leg-val}

In this section, we provide Table~\ref{tab:leg-val}, a two-by-two table of examples of gender data and gender imputation used for anti-discrimination purposes, for additional clarity. All four example measurements center around measuring different kinds of gender disparities in the employment setting.

\begin{table*}[tb]
\centering
\begin{tabular}{|C{2cm}|C{7cm}|C{7cm}|}
\hline
 & \textbf{Valid} & \textbf{Invalid} \\
\hline
\textbf{Legitimate} & using self-reported gender identity to measure wage differences & using self-reported gender identity to measure online recruiter discrimination \\
\hline
\textbf{Illegitimate} & imputing perceived gender from scraped online photos to measure online recruiter discrimination & imputing perceived gender from scraped online photos to measure wage differences \\
\hline
\end{tabular}
\caption{Examples of different disparity measurements using gender and imputed gender. Imputed values are always illegitimate, while properly collected self-reported data is one example of legitimate data. The valid examples use aspects of gender aligned with the form of discrimination we intend to measure (gender identity corresponding to the gender wage gap, and perceived gender corresponding to treatment by recruiters perceiving candidates), while invalid ones do not. Note that self-reported gender identity, while legitimate, is not the basis on which online gender discrimination may occur, and may lead to less valid measurements.}
\label{tab:leg-val}
\end{table*}

\end{document}